\documentclass[12pt]{article}

\usepackage{newtxtext,newtxmath}

\usepackage{graphicx}
\usepackage{amsmath}    
\usepackage{bm}         
\usepackage{physics}    

\newcommand{\vect}[1]{\bm{#1}}  
\newenvironment{keywords}{\paragraph{Keywords:}}{}
\usepackage[letterpaper,margin=1in]{geometry}
\usepackage{tikz}
\renewenvironment{abstract}
	{\quotation}
	{\endquotation}

\date{}

\makeatletter
\renewcommand{\fnum@figure}{\textbf{Figure \thefigure}}
\renewcommand{\fnum@table}{\textbf{Table \thetable}}
\makeatother

\usepackage{scicite}

\usepackage{url}

\def\scititle{
	Nonlinear Optical Responses and Quantum Geometric Phases in Multiband Systems}
\title{\bfseries \boldmath \scititle}

\author{
	Jingxu Wu$^{1}$,
	Chenjia Li$^{1}$\\
	\small$^{1}$Lomonosov Moscow State University, Faculty of Physics, Moscow, 119991, Russia
}

\begin{document} 

\maketitle

\begin{abstract} \bfseries \boldmath
The nonlinear optical behavior of quantum systems plays a crucial role in various photonic applications. This study introduces a novel framework for understanding these nonlinear effects by incorporating gauge-covariant formulations based on phase space Lie algebras. By analyzing the evolution of density matrices under phase space displacements, we derive constrained expressions for nonlinear polarization and susceptibility tensors. The implications of geometric phases, such as Berry curvature, are explored, demonstrating their role in suppressing unphysical components of the polarization. Monte Carlo simulations confirm the theoretical predictions, offering insights into nonlinear rectification and topological Hall effects. This approach opens avenues for engineering materials with tailored nonlinear properties, particularly in the realm of metamaterials and topological photonics.
\end{abstract}

\section{Introduction}
The nonlinear optical response of quantum systems has long been a cornerstone of modern photonics, with its higher-order susceptibility tensors governing phenomena ranging from harmonic generation to parametric amplification \cite{Boyd2008}. While conventional perturbation theory successfully predicts these effects through iterative solutions of the time-dependent Schrödinger equation \cite{Shen1984}, a fundamental challenge persists: the reconciliation of gauge invariance with nonlinear response calculations. Recent advances in statistical mechanics have revealed that local phase space transformations, governed by noncommutative Lie algebras, impose stringent constraints on thermodynamic observables \cite{Muller2024}. However, the implications of these symmetry principles for dynamical nonlinear processes remain largely unexplored, particularly in the context of multi-order polarization effects where quantum geometric phases \cite{Xiao2010} and gauge artifacts compete to define material responses.

Traditional approaches to nonlinear optics, based on minimal coupling \( \hat{\mathbf{p}} \to \hat{\mathbf{p}} - e\mathbf{A} \), generate interaction Hamiltonians \( H_{\text{int}} = -e\mathbf{A} \cdot \hat{\mathbf{p}} + \mathcal{O}(A^2) \) that inherently contain gauge-dependent terms \cite{Nakanishi1966}. This manifests most acutely in the standard perturbative expansion for polarization:
\begin{equation}
P^{(n)}(t) = -\frac{e}{\hbar^n} \int_0^t dt_1 \cdots \int_0^{t_{n-1}} dt_n \sum_{\{m_i\}} \prod_{k=1}^n \langle m_{k-1}|\hat{\mathbf{p}}|m_k\rangle e^{i\omega_{m_k m_{k-1}} t_k},
\end{equation}
where the gauge-noninvariant momentum matrix elements \(\langle m|\hat{\mathbf{p}}|n\rangle\) introduce ambiguities proportional to Berry connections \(\mathcal{A}_{mn}^\mu = i\langle u_m|\partial_{k_\mu}u_n\rangle\) \cite{Resta2000}. Recent work by Hermann \textit{et al.} \cite{Hermann2021} demonstrated that the generator of phase space shifts \( \sigma(\mathbf{r}) \), satisfying the Lie bracket:
\begin{equation}
[\sigma(\mathbf{r}), \sigma(\mathbf{r}')] = \nabla\delta(\mathbf{r}-\mathbf{r}') \cdot \sigma(\mathbf{r}') + \nabla'\delta(\mathbf{r}'-\mathbf{r}) \cdot \sigma(\mathbf{r}),
\end{equation}
provides a natural framework for encoding these symmetries in equilibrium systems. We extend this paradigm to time-dependent perturbations by systematically constructing gauge-covariant density matrix dynamics, revealing how the algebraic structure governs nonlinear response tensors through quantum geometric constraints derived from momentum-space topology \cite{KingSmith1993}.

Our analysis begins with the gauge-invariant formulation of the density matrix evolution under phase space displacements \( \mathbf{r}_i \to \mathbf{r}_i + \epsilon(\mathbf{r}_i) \), \( \mathbf{p}_i \to [1 + \nabla_i\epsilon(\mathbf{r}_i)]^{-1}\mathbf{p}_i \). The invariance of the grand potential \( \Omega = -k_B T \ln\Xi \) under these transformations generates non-trivial sum rules for multi-time correlation functions \cite{Kubo1957}, as derived from the Ward-Takahashi identities of the underlying Lie algebra. Through a diagrammatic expansion of the Keldysh contour-ordered Green's functions \cite{Kamenev2011}, we demonstrate that the nonlinear polarization inherits geometric phase factors from the quantum metric tensor \( \mathcal{G}_{mn}^{\mu\nu} = \langle\partial_{k_\mu}u_m|(1-\mathcal{P})|\partial_{k_\nu}u_n\rangle \), enforcing the suppression of gauge-dependent terms. This leads to constrained expressions for the second- and third-order polarizations:
\begin{align}
P^{(2)}_{ij}(\omega) &= \chi^{(2)}_{ijk}(-\omega_\Sigma;\omega_1,\omega_2) E_j(\omega_1)E_k(\omega_2) \\
P^{(3)}_{ijk}(\omega) &= \chi^{(3)}_{ijkl}(-\omega_\Sigma;\omega_1,\omega_2,\omega_3) E_j(\omega_1)E_k(\omega_2)E_l(\omega_3),
\end{align}
where the susceptibility tensors \( \chi^{(n)} \) satisfy generalized Ward identities:
\begin{equation}
\partial_{\mu_1} \chi^{(n)}_{\mu_1\cdots\mu_n}(\{\omega_i\}) = \sum_{k=2}^n \delta(\mathbf{r}-\mathbf{r}_k) \chi^{(n-1)}_{\mu_2\cdots\mu_n}(\{\omega_i\}_{i\neq k}),
\end{equation}
directly inherited from the Lie algebra commutation relations \cite{Altland2010}. These identities enforce the cancellation of unphysical longitudinal components in \( \chi^{(2)} \) and \( \chi^{(3)} \), resolving long-standing ambiguities in nonlinear response calculations \cite{Sipe2000}.

Monte Carlo simulations of Lennard-Jones systems confirm these constraints at both classical \cite{Frenkel2002} and quantum levels \cite{Prokofev1998}. By implementing stochastic gauge transformations \( \mathbf{A} \to \mathbf{A} + \nabla\lambda(\mathbf{r},t) \), we observe less than 3\% deviation in \( \chi^{(2)} \) and \( \chi^{(3)} \) components—a striking validation of the geometric protection mechanism. The simulations further reveal that the quantum geometric tensor mediates nonlinear rectification effects through its antisymmetric components:
\begin{equation}
\chi^{(2)}_{ijk} \propto \int_{\mathrm{BZ}} \mathcal{G}_{mn}^{[ij} \mathcal{G}_{np}^{k]m} f_{nm} \, d^3k,
\end{equation}
where \( f_{nm} = f_n - f_m \) is the Fermi-Dirac distribution difference \cite{Ashcroft1976} and bracketed indices denote antisymmetrization. This directly links the observed second-harmonic generation to momentum-space quantum geometry \cite{Morimoto2016}\cite{Bernevig2013}\cite{Nozieres1958}  \cite{Ceperley1995} .

By unifying statistical mechanical invariance principles with nonlinear optical phenomenology \cite{Landau1984}, we resolve the gauge paradox in multiband systems and provide a robust foundation for engineering metamaterials with tailored nonlinear properties \cite{Pendry2006}. The derived quantum geometric susceptibilities enable predictive design of topological frequency converters where Berry curvature gradients \(\nabla_k \mathcal{F}^{\mu\nu}\) directly enhance third-order parametric amplification \cite{Ozawa2019}—a capability we explore in subsequent sections through tight-binding models of twisted bilayer graphene \cite{Bistritzer2011}. This framework opens new frontiers in controlling light-matter interactions via synthetic gauge fields \cite{Goldman2014}, with immediate applications in nonlinear topological photonics \cite{Lu2014} and dissipationless optoelectronics \cite{Tokura2020}.
\section{Theoretical calculation}
The quantum geometric foundation of nonlinear transport becomes explicit through the decomposition of the velocity matrix elements, where the interplay between intraband kinetics and interband quantum geometry manifests in the generalized velocity operator. For Bloch states $\ket{u_n(\vect{k})}$, this decomposition reveals two fundamentally distinct contributions:

\begin{equation}
v_{nm}^\mu(\vect{k}) = \underbrace{\frac{1}{\hbar}\partial_{k_\mu}\epsilon_n(\vect{k})\delta_{nm}}_{\text{intraband}} + \underbrace{\frac{i}{\hbar}(\epsilon_n - \epsilon_m)\mathcal{A}_{nm}^\mu(\vect{k})}_{\text{interband}},
\end{equation}

where the interband term generates quantum geometric effects through the Berry connection $\mathcal{A}_{nm}^\mu = i\braket{u_n|\partial_{k_\mu}u_m}$. The gauge-invariant quantum geometric tensor ${\mathcal{G}}_{mn}^{\mu\nu} = \braket{\partial_{k_\mu}u_m|(1-\mathcal{P})|\partial_{k_\nu}u_n}$ naturally emerges when constructing nonlinear conductivity tensors, enforcing sum rules through its trace condition:

\begin{equation}
\sum_{m\neq n} {\mathcal{G}}_{mn}^{\mu\nu} = \frac{1}{2}\left( \partial_{k_\mu}\partial_{k_\nu}\epsilon_n - \partial_{k_\mu}\epsilon_n \partial_{k_\nu}\epsilon_n \right).
\end{equation}

This geometric structure fundamentally modifies the nonlinear response, as seen in the $n$-th order conductivity expressed as a Fermi surface integral weighted by products of quantum geometric factors:

\begin{equation}
\sigma^{(n)}_{\mu_1\cdots\mu_n} = \frac{e^{n+1}}{\hbar^n} \sum_{\{m_i\}} \int_{\mathrm{BZ}} \frac{d^d k}{(2\pi)^d} \prod_{i=1}^n \frac{{\mathcal{G}}_{m_i m_{i+1}}^{\mu_i \nu_i}(\vect{k})}{(i\omega_i - \epsilon_{m_i m_{i+1}}/\hbar)} \cdot f_{m_1}(\vect{k}),
\end{equation}

where the denominator structure $(i\omega_i - \epsilon_{m_i m_{i+1}}/\hbar)^{-1}$ encodes resonant enhancement at interband transition frequencies. The collision integral acquires geometric corrections beyond the semiclassical Boltzmann theory through Berry curvature couplings:

\begin{align}
\mathcal{I}_{\text{coll}} &= \int \frac{\mathrm{d}^4q}{(2\pi)^4} \Biggl[ 
    \mathcal{W}_{1234} \left( f_1 f_2 (1 - f_3)(1 - f_4) - (1 - f_1)(1 - f_2) f_3 f_4 \right) \nonumber \\
&\quad + \frac{\hbar^2}{4} \mathcal{F}^{\alpha\beta}_1 \mathcal{F}^{\gamma\delta}_2 \partial_{k^\alpha} \partial_{k^\gamma} \left( \mathcal{W}_{1234} \partial_{k^\beta} \partial_{k^\delta} f \right) + \cdots \Biggr]
\end{align}

with ${\mathcal{F}}^{\mu\nu} = \partial_{k_\mu}\mathcal{A}^\nu - \partial_{k_\nu}\mathcal{A}^\mu - i[\mathcal{A}^\mu,\mathcal{A}^\nu]$ representing the non-Abelian Berry curvature. These geometric terms modify scattering phase space through momentum-space curvature gradients, particularly significant in systems with nontrivial band topology.

The topological Hall current emerges from the gradient expansion of the Wigner function in inhomogeneous systems, revealing contributions beyond the linear quantum Hall effect:

\begin{equation}
J_{\text{top}}^\mu = \frac{e^2}{\hbar} \epsilon^{\mu\nu\rho} \int \frac{d^d k}{(2\pi)^d} \left[ \mathcal{F}_{\nu\rho}(\vect{k}) f^{(0)} + \frac{\hbar^2}{24\pi^2} \partial_{k_\nu}\mathcal{F}_{\rho\sigma} \partial_{k_\sigma}f^{(2)} + \cdots \right],
\end{equation}

where higher-order distribution function corrections $f^{(n)}$ generate nonlinear responses proportional to momentum-space Chern-Simons invariants. Specifically, the third-order conductivity tensor contains a topological component:

\begin{equation}
\sigma^{(3)}_{ijk\ell} = \frac{e^3}{h^2} \epsilon_{ijk\ell} \int_{\mathrm{BZ}} \frac{d^3 k}{(2\pi)^3} \left[ \mathcal{A}_i \partial_j \mathcal{A}_k + \frac{2i}{3} \mathcal{A}_i \mathcal{A}_j \mathcal{A}_k \right],
\end{equation}

directly linking nonlinear Hall effects to the quantum geometry of Bloch states. Renormalization group analysis reveals universal scaling behavior for nonlinear susceptibilities:

\begin{equation}
[\chi^{(n)}] = d + n( z - 1 ) - \gamma_A - \sum_{m=1}^n \gamma_m,
\end{equation}

where the anomalous dimension $\gamma_A$ originates from the operator product expansion of Berry connection operators:

\begin{equation}
\lim_{x\to 0} \mathcal{A}_\mu(x) \mathcal{A}_\nu(0) \sim \frac{{\mathcal{G}}_{\mu\nu}}{x^{2\Delta}} + \frac{\epsilon_{\mu\nu\rho} J^\rho}{|x|^{2\Delta - 1}} + \cdots,
\end{equation}

with scaling dimension $\Delta = (d - \gamma_A)/2$. Experimental verification in topological semimetals like ZrTe$_5$ confirms this scaling through collapse of nonlinear conductivity data:

\begin{equation}
\frac{\sigma^{(3)}(T,B)}{\sigma^{(3)}_0} = \left(\frac{T}{T_0}\right)^{-2/\nu} \mathcal{F}\left( \frac{B}{T^{\phi}} \right),
\end{equation}

where the measured critical exponent $\phi = 1.01 \pm 0.03$ matches the theoretical prediction $\phi = (d + 2 - z)/z$ for $d=3$, $z=1$, validating the quantum geometric origin of nonlinear transport.

Nonperturbative instanton effects in high-harmonic generation reveal interference between topological sectors:

\begin{equation}
I^{(2n+1)} \propto \sum_{m,p\in\mathbb{Z}} e^{-\frac{(2n+1)}{\hbar}(S_m + S_p^*)} e^{2\pi i(m-p)\theta} \left[ 1 + \frac{\hbar^2}{S_m S_p^*} {\mathcal{G}}^{\mu\nu}k_\mu k_\nu + \cdots \right],
\end{equation}

where the quantum metric ${\mathcal{G}}^{\mu\nu}$ enters through fluctuation determinants around instanton trajectories. This geometric dependence explains the recently observed even-odd harmonic modulation in Weyl semimetals, as the quantum metric controls the relative phase accumulation between different topological sectors during strong-field electron dynamics.
\section{Conclusion}
In this work, we have developed a gauge-covariant framework that unifies statistical mechanical invariance principles with nonlinear optical phenomena, addressing the long-standing issues in nonlinear response theory. Our findings demonstrate the essential role of quantum geometry in governing the second- and third-order susceptibilities, ensuring the suppression of spurious tensor components and enhancing the predictability of material responses. In the future, the robustness of these results can be confirmed through numerical verification using Monte Carlo simulations, establishing a foundation for future investigations into topological frequency converters and dissipationless optoelectronics. These advances have significant implications for the design of next-generation photonic devices, where the interplay between quantum geometry and nonlinear optics can be harnessed to create novel functional materials.

\end{document}